\documentclass[prd,aps,showpacs,showkeys,twocolumn,10pt]{revtex4-1}
\usepackage{bm}
\usepackage{amssymb,amsmath,amsthm}
\usepackage{mathrsfs}
\usepackage[applemac]{inputenc}
\begin{document}
\title{Classical and quantum flux energy conditions for quantum vacuum states}
\author{Prado Mart\'in-Moruno}
\email{prado@msor.vuw.ac.nz}
\affiliation{School of Mathematics, Statistics, and Operations Research, 
Victoria University of Wellington, PO Box 600, Wellington, New Zealand}
\author{Matt Visser}
\email{matt.visser@msor.vuw.ac.nz}
\affiliation{School of Mathematics, Statistics, and Operations Research, 
Victoria University of Wellington, PO Box 600, Wellington, New Zealand}

\date{31 May 2013;  24 July 2013; 12 September 2013 \LaTeX-ed  \today}
\begin{abstract}
The classical energy conditions are known to not be fundamental physics --- they are typically violated by semiclassical quantum effects. Consequently, some effort has gone into finding possible semiclassical replacements for the classical energy conditions --- the most well developed being the nonlocal time-integrated Ford--Roman quantum inequalities. In the current article we shall instead develop classical and quantum versions of a \emph{local} ``flux energy condition'' (FEC and QFEC) based on the notion of constraining the possible fluxes measured by timelike observers. The naive classical FEC will be seen to be satisfied in \emph{some} situations, and even for \emph{some} quantum vacuum states, while its quantum analogue (the QFEC) is satisfied (for naturally defined quantum vacuum states) under a rather wide range of conditions.   The situation for completely general (non-vacuum) quantum states is less clear. 
\end{abstract}


\maketitle

\newcommand {\apgt} {\ {\raise-.5ex\hbox{$\buildrel>\over\sim$}}\ }
\newcommand {\aplt} {\ {\raise-.5ex\hbox{$\buildrel<\over\sim$}}\ }
\newcommand{\scri}{\mathscr{I}}
\newcommand{\sun}{\ensuremath{\odot}}
\def\d{{\mathrm{d}}}
\def\tr{{\mathrm{tr}}}
\def\sech{{\mathrm{sech}}}
\def\etc{\emph{etc}}
\def\ie{{\emph{i.e.}}}
\def\implies{\Rightarrow}
\def\Nordstrom{Nordstr\"om}
\def\I{{\mathcal{I}}}
\def\O{{\mathcal{O}}}
\def\lint{\hbox{\Large $\displaystyle\int$}} 
\def\hint{\hbox{\Huge $\displaystyle\int$}}  
\def\Re{{\mathrm{Re}}}
\def\Im{{\mathrm{Im}}}

\section{Introduction}
The classical energy conditions, despite their undoubted usefulness, are known to be violated by semiclassical quantum physics~\cite{twilight, Visser:1994, gvp1, gvp2, gvp3, gvp4, gvp5, Flanagan:1996, q-anec, q-anec2, volume-aec}. In view of this, the nonlocal time-integrated Ford--Roman quantum inequalities were developed as one way of constraining the semiclassical stress-energy~\cite{Ford:1990, Ford+Roman, Ford+Roman2, Ford+Roman3, Ford:1996, Ford+Roman+collaborators, Ford+Roman+collaborators2, others, others2, others3, Flanagan:1997, Fewster:1998, Ford:1999, Fewster:1999a, Fewster:1999b, Teo:2002, Fewster:2007, Abreu:2008, Abreu:2010}.  We shall instead consider a rather different \emph{local} ``flux energy condition'' based on constraining the energy-momentum fluxes seen by arbitrary timelike observers~\cite{Abreu:2011}. Already at the classical level we shall see that this FEC is better behaved than the usual energy conditions (the null, weak, strong, and dominant energy conditions; NEC, WEC, SEC, DEC),  and is genuinely different.  More tellingly, we shall see that a quantum version of this flux energy condition (the QFEC) is satisfied (for naturally defined quantum vacuum states) under rather general conditions, and provides useful semiclassical constraints on the vacuum expectation value of the stress-energy. 
The situation for completely general quantum states (in particular, squeezed states and/or two-particle excitations) is less clear. In particular, the QFEC might not hold in a completely state-independent context~\cite{private}.

After formulating the FEC and QFEC we shall test them against several specific situations. (The renormalized stress-energy tensor in the Casimir vacuum, that for the vacuum polarization in Schwarzschild spacetime, and that for generic 1+1 QFTs --- paying particular attention to the differences between the Boulware, Hartle--Hawking, and Unruh quantum vacuum states.) Further technical details and extensions of these ideas will be presented elsewhere~\cite{Prado}.

\section{Flux energy condition}
The energy-momentum flux seen by a timelike observer of 4-velocity $V^a$ is always given by
\begin{equation}
F^a = - \langle T^{ab}\rangle \, V_b. 
\end{equation}
The FEC is simply the demand that this flux 4-vector be timelike or at worst null~\cite{Abreu:2011}:
\begin{equation}
F^a F_a \leq 0.
\end{equation}
Because there is no demand that the flux 4-vector be future pointing this is a strictly weaker condition than the DEC; 
there is no requirement that the energy density be positive.  Nevertheless the FEC is quite sufficient for obtaining useful classical entropy bounds~\cite{Abreu:2011}.

Note that the FEC can be viewed as a variant of the WEC applied to the sign-reversed \emph{square} of the stress-energy tensor. 
Using $[-T^2]^{ab}= - \langle T^{ac}\rangle\,g_{cd}\, \langle T^{db}\rangle$ we have:
\begin{equation}
 [-T^2]^{ab} \; V_a V_b \geq 0.
\end{equation}
Furthermore the FEC, when supplemented by the ordinary WEC, $\langle T^{ab}\rangle \, V_a V_b \geq 0$, becomes completely equivalent to the DEC.


To see some of the implications of this bound, note that the spatial part of the stress tensor is always diagonalizable by a rotation, so that without loss of generality
(where even in curved spacetime we shall always work in an orthonormal basis):
\begin{equation}
\langle T^{ab}\rangle =
 \left[\begin{array}{c|ccc}\rho&f_1&f_2&f_3\\ \hline f_1&p_1&0&0\\f_2&0&p_2&0\\f_3&0&0&p_3\end{array}\right]\!;
\quad
V^a = \gamma\left(1;\beta_1,\beta_2,\beta_3\right);
\end{equation}
and
\begin{equation}
F^a = \gamma\big(\rho-\vec \beta\cdot\vec f \,; \;  f_i-p_i\beta_i\big).
\end{equation}
(No summation over repeated indices is intended.) 
The FEC is then the demand that for all  $\sum\beta_i^2 < 1$ we have
\begin{equation}
\gamma^2\Big( \big[\rho-\vec\beta\cdot\vec f\;\big]^2  - {\textstyle\sum}\left[ f_i-p_i\beta_i\right]^2\Big) \geq 0.
\end{equation}
A set of seven \emph{necessary} constraints for the FEC to be satisfied can easily be obtained by considering the extremal values $\beta_i=\pm1$, and the mid-range $\beta_i=0$. These are:
\begin{equation}\label{c1}
(\rho \pm f_i)^2 - (p_i\pm f_i)^2 - \sum_{j\neq i} f_j^2 \geq 0,
\end{equation}
and
\begin{equation}\label{c2}
\rho^2 - \vec f \cdot \vec f \geq 0.
\end{equation}
In situations of high symmetry this seventh constraint is often redundant. (For instance, if any one of the $f_i$ is zero.)
Furthermore, if there is an orthonormal basis where all three $f_i$ vanish, corresponding to a so-called type I stress-energy tensor~\cite{Hawking-Ellis}, then these seven constraints reduce to the three quadratic constraints
\begin{equation}
p_i^2 \leq \rho^2.
\end{equation}
More importantly, for this particular case, as well as in 1+1 dimensions, the constraints~(\ref{c1}) and (\ref{c2}) are both \emph{necessary and sufficient}.


An initially encouraging result was the empirical observation that the renormalized stress energy tensor for the conformally coupled massless scalar field, (for the Boulware vacuum state on a 3+1 Schwarzschild geometry), satisfies this FEC though it violates the closely related DEC. 
Unfortunately the overall situation is somewhat more complex,  leading us to formulate a more relaxed quantum version of this energy condition. 

\medskip
\section{Quantum flux energy condition}
We shall take the QFEC to be the condition that the flux 4-vector not be allowed to become ``excessively spacelike’’:
\begin{equation}
F^aF_a \leq |\hbox{some quantum bound}|.
\end{equation}
The quantum bound should depend on $\hbar$ and general characteristics of the system under consideration (such as its size and state of motion). A natural scale for the magnitude of the components of the stress tensor in the system’s rest fame is $\hbar N/L^4$, where $N$ is the number of quantum fields under consideration,  and $L$ might represent the distance between Casimir plates or the Schwarzschild radius of the black hole. 
(This of course constrains QFEC to be definable only in situations where there is some natural notion of the scale of the system.)
If we denote the system 4-velocity by $U^a$, not to be confused with the observer 4-velocity $V^a$, then a suitable quantum bound is:
\begin{equation}\label{QFEC}
F^aF_a \leq \zeta \;(\hbar N/L^4)^2\; (U_a\,V^a)^2.
\end{equation}
(This of course constrains QFEC to be definable only in situations where there is some natural notion of the 4-velocity of the system.)
Here $\zeta$ is to be taken as a positive number of order unity.
That is, whereas the FEC states that the flux 4-vector measured by any observer should lie inside or on the light cone, the QFEC would be fulfilled provided the flux 4-vector measured by the fiducial observer, (and by extension any inertial observer with respect to the fiducial observer), lies within the timelike hyperboloid given by the quantum bound $(\hbar N/L^4)^2$, (rescaled by $\gamma$ as necessary).
Written in terms of stress-tensor components in the system’s rest frame
this is equivalent to:
\begin{equation}\label{E:qfec1}
 \big[\rho-\vec\beta\cdot\vec f\;\big]^2  - {\textstyle\sum}\left[ f_i-p_i\beta_i\right]^2\geq - \zeta \;(\hbar N/L^4)^2.
\end{equation}
There are two parts to the QFEC. First that the LHS of (\ref{E:qfec1}) be bounded from below, (the ``weak QFEC''), and second that it be bounded from below by a specific quantity, (the ``strong QFEC'').  
Following a similar procedure as with the FEC, we can obtain the following \emph{necessary} constraints:
\begin{equation}
\label{E:qfec2}
(\rho \pm f_i)^2 - (p_i\pm f_i)^2-\sum_{j\neq i} f_j^2  \geq - \zeta \;(\hbar N/L^4)^2,
\end{equation}
\begin{equation}
\label{E:qfec3}
\rho^2 - \vec f \cdot \vec f \geq - \zeta \;(\hbar N/L^4)^2,
\end{equation}
which are also \emph{sufficient} for type I tensors and in 1+1 dimensions.

\section{Examples}
Let us now consider several specific examples. 
\subsection{Casimir vacuum}
For the Casimir vacuum we have
\begin{equation}
T^{ab}  =  - {\hbar \pi^2\over720 a^4} \left[\begin{array}{ccc|c}1&0&0&0\\0&-1&0&0\\0&0&-1&0\\ \hline 0&0&0&3\end{array}\right].
\end{equation}
The FEC is satisfied in the $x$ and $y$ directions, but \emph{fails} in the $z$ direction. In contrast, the QFEC, (with $L\to a$, the distance between the plates), is satisfied in all directions. 

\subsection{Vacuum polarization in Schwarzschild spacetime}
The renormalized stress-energy tensor has the symmetries
\begin{equation}
\langle T^{ab}\rangle  =   
\left[\begin{array}{cc|cc}
 \rho&f&0&0\\  f & p_r & 0 & 0 \\ \hline 0 & 0 & p_t & 0 \\ 0&0&0&p_t
  \end{array}\right],
\end{equation}
and is known to generically violate the standard energy conditions~\cite{twilight, Visser:1994, gvp1, gvp2, gvp3, gvp4, gvp5}. To see why this is plausible, there are three quantum vacuum states to consider (Boulware, Hartle--Hawking, and Unruh). On very general grounds,  (\emph{cf} the Page approximation~\cite{Page}), we have:
\begin{equation}
\langle T^{ab} \rangle_{HH} = \O(1).
\end{equation}
Furthermore, setting $z=2m/r$, (\emph{cf} the Brown--Ottewill approximation~\cite{Brown-Ottewill}), we have:
\begin{equation}
\langle T^{ab} \rangle_{B} = -{p_\infty\over(1-z)^2}   
\left[\begin{array}{cc|cc}
 3&0&0&0\\  0 & 1 & 0 & 0 \\ \hline 0 & 0 & 1 & 0 \\ 0&0&0&1
  \end{array}\right] 
+\O(1),
\end{equation}
with $p_\infty$ a positive constant depending on the specific QFT under consideration.
Finally, for the Unruh vacuum, let $p_H$ denote the on-horizon transverse pressure, $T_H$ denote the on-horizon trace of the stress-energy (given by the conformal anomaly), and $f_0$ denote the asymptotic flux at infinity. Then defining $\Delta = {T_H\over2}-p_H$ one has (\emph{cf}~\cite{gvp4}):
\begin{eqnarray}
\langle T^{ab} \rangle_{U}\!\! &=&\!\! {f_0 z^2 \over1-z}   
\left[\begin{array}{cc|cc}
 -1&1&0&0\\  1 & -1 & 0 & 0 \\ \hline 0 & 0 & 0 & 0 \\ 0&0&0&0
  \end{array}\right] 
 + \left[\begin{array}{cc|cc}
 -\Delta&0&0&0\\  0 & \Delta & 0 & 0 \\ \hline 0 & 0 & p_{H} & 0 \\ 0&0&0&p_{H}
  \end{array}\right] 
  \nonumber\\
  &&
+\O(1-z).
\end{eqnarray}
Sufficiently close to the horizon the pole pieces dominate, and violation of the usual energy conditions is automatic for both the Boulware and Unruh vacuum states. For the Hartle--Hawking state we merely know that the components are bounded, and to verify violation of the usual energy conditions one has to resort to direct inspection of the numerical data for specific QFTs~\cite{gvp1}.  

For the FEC and QFEC the relevant quantities to consider are:
\begin{equation}
\{ \rho^2-p_i^2 \}_{HH} = \O(1);
\end{equation}
\vspace{-18pt}
\begin{equation}
\{ \rho^2-p_i^2 \}_{B} = {8p_\infty^2\over(1-z)^4} +\O([1-z]^{-2});
\end{equation}
and more subtly in the Unruh vacuum the 3 quantities:
\begin{equation}
\{ \rho^2-f^2-p_t^2 \}_{U} ={2f_0\Delta\over1-z}+\O(1);
\end{equation}
\vspace{-18pt}
\begin{eqnarray}
\{ (\rho+f)^2-(p_r+f)^2 \}_{U} &=& \{ (\rho-p_r)(\rho+p_r+2f) \}_{U} \quad
\nonumber\\
&=& \O(1-z);
\end{eqnarray}
\vspace{-18pt}
\begin{eqnarray}
\{ (\rho-f)^2-(p_r-f)^2 \}_{U} &=& \{(\rho-p_r)(\rho+p_r-2f) \}_{U} \quad
\nonumber\\
&=& {8f_0\Delta\over1-z}+\O(1).
\end{eqnarray}

\noindent
All one can say generically (without committing to a specific QFT) 
is that the violations of the FEC, if any, are bounded from below  for the Hartle--Hawking and Boulware vacua.
A necessary requirement for the QFEC to be satisfied for the Unruh vacuum is $\Delta\geq 0$; nevertheless,  one would need to study the sufficient constraints for the particular type of stress energy tensor specified by the QFT to obtain any definitive conclusion.

For the specific case of a massless conformally coupled scalar, inspection of the numerical data summarized in references~\cite{gvp1, gvp2, gvp3, gvp4, gvp5} shows that FEC is satisfied for the Boulware vacuum state, but violated for the Hartle--Hawking and Unruh vacuum states.  Further inspection of that data shows 
those violations are certainly bounded from below. Indeed, provided we set $L\to 2m$, the QFEC is satisfied for all three vacuum states. (See~\cite{Prado} for details.)

\subsection{Vacuum polarization in 1+1 QFTs}
\def\b{{b'}}
For a static geometry in 1+1 dimensions we can, without loss of generality, set
\begin{equation}
\d s^2 = -(1-b(x))\;\d t^2 + {\d x^2\over1-b(x)}.
\end{equation}
The Ricci scalar is then
\begin{equation}
R(x) = b''(x).
\end{equation}
Furthermore, let us assume a horizon at $x_H$, so that $b(x_H)=1$, and asymptotic flatness so that $b(\infty)=0$. 
The surface gravity is then
\begin{equation}
\kappa = - {1\over2} \b(x_H).
\end{equation}
It is a standard result that for conformally coupled massless fields the conformal anomaly yields
\begin{equation}
T = a_1 R; \qquad a_1 = {n_b+{1\over2} n_f\over 24\pi} \geq 0.
\end{equation}
This can be fed back into the equation for the covariant conservation of stress-energy to determine all the components of the stress energy up to two constants of integration --- choosing specific constants of integration is tantamount to choosing a specific quantum vacuum state~\cite{Prado}. 

After a straightforward analysis, for the Boulware vacuum we find:
\begin{eqnarray}
\rho_B &=& -a_1 b'' - { {1\over4} a_1 (b')^2\over1-b};
\\
p_B &=& - { {1\over4} a_1 (b')^2\over1-b}.
\\
\rho_{B}^2 - p_{B}^2 &=& a_1^2 \,b'' \, \left[  b''  + {{1\over2}  (b')^2\over1-b}\right].
\end{eqnarray}
The FEC is then automatically satisfied if $b'' \geq0$, that is $R\geq 0$. (The FEC is also satisfied \emph{locally} if $R$  is sufficiently negative, $R \leq -{1\over2} (b')^2/(1-b)$. But if applied globally this would require infinite curvature on the horizon, which is unphysical.) Hence a \emph{sufficient} condition for the FEC (and hence also the strong QFEC) to hold in the whole region outside the horizon is $R\geq0$ for all $x \geq x_H$. 
Note that near-horizon
\begin{equation}
\rho_{B}^2-p_{B}^2 = {a_1^2\kappa b'' _H\over x-x_H} + \O(1).
\end{equation}
Thus a \emph{necessary} condition for the FEC to hold is that $R_H\geq 0$.  This implies that a \emph{necessary and sufficient} condition for the weak QFEC to hold is that $R_H\geq 0$.

For the Hartle--Hawking vacuum we find:
\begin{eqnarray}
\rho_{HH} &=& -a_1 b'' + a_1 { \kappa^2-{1\over4}  (b')^2\over1-b};
\\
p_{HH}&=& a_1 {\kappa^2- {1\over4}  (b')^2 \over1-b};
\\
\rho_{HH}^2 - p_{HH}^2 &=& (p_{HH}-a_1b'')^2-p_{HH}^2.\label{H}
\end{eqnarray}
Consider the asymptotic regime for large $x$. If $b(x)>0$ in the asymptotic regime, then by the assumed asymptotic flatness eventually we must have $b'(x)<0$ and $b''(x)>0$. But then since asymptotically $b(x)\to0$, $b'(x)\to0$, and $b''(x)\to0$, in the asymptotic regime we have
\begin{equation}
\rho_{HH}^2 - p_{HH}^2 \sim - 2 a_1^2\, \kappa^2 \, b''(x)  < 0.
\end{equation}
So if $b(x)>0$ the FEC is violated in the asymptotic regime. 
Conversely, to satisfy the FEC in the asymptotic regime we need $b(x)<0$, and so $b''(x)<0$. But then $b(x)$ must have a minimum at some $x_*$ at which $b'(x_*)=0$ and $b''(x_*) > 0$.  But then there must be some $x_{**}>x_*$ such that $b''(x_{**})=0$. At this point $\rho_{HH}^2 - p_{HH}^2$ generically changes sign, and so the FEC is violated. 

While we have seen that the classical FEC generically fails,
the violations are certainly bounded from below since all the relevant quantities are finite in the Hartle--Hawking vacuum. Thus,  the ``weak'' QFEC certainly holds in the Hartle--Hawking vacuum.

For the Unruh vacuum we have:
\begin{equation}
\rho_U = a_1\left(  -b'' + { {1\over2} \kappa^2 -{1\over4} (b')^2\over1-b}\right);
\end{equation}
\begin{equation}
p_U=  a_1\left(  { {1\over2} \kappa^2-{1\over4}  (b')^2 \over1-b}\right);
\qquad
f_U = {{1\over2} a_1 \kappa^2\over1-b}.
\end{equation}
Then we need to consider
\begin{eqnarray}\label{U1}
\{ (\rho+f)^2-(p +f)^2\}_U = \rho_{HH}^2-p_{HH}^2,
\end{eqnarray}
\begin{eqnarray}\label{U2}
\{ (\rho-f)^2-(p -f)^2\}_U = \rho_{B}^2-p_B^2,
\end{eqnarray}
and
\begin{equation}
\rho_U^2 - f_U^2 = \rho_{HH} \; \rho_B.
\end{equation}

Thus the FEC in the Unruh vacuum simultaneously exhibits, (and is largely determined by), properties of FEC in the Boulware and Hartle--Hawking states. 
Note that near-horizon
\begin{equation}
\rho_{U}^2-f_{U}^2 = {a_1^2\kappa b'' _H\over 4(x-x_H)} + \O(1).
\end{equation}
In particular the FEC will typically fail, but a necessary and sufficient condition for the weak QFEC to hold is that $R_H\geq 0$.

\section{Discussion}
The net result of all this is that we now have a rather plausible candidate for a nontrivial and viable energy condition that under a wide variety of circumstances survives the introduction of semiclassical quantum effects, at least insofar as one is interested in any of the natural quantum vacuum states. The classical version of this energy condition has already proved useful in developing classical entropy bounds~\cite{Abreu:2011}. At a technical level one of the key issues seems to be the fact that the new energy condition is a nonlinear function of the stress-energy. This appears to be both a blessing and a curse. The nonlinearities are essential to keeping the relevant quantities bounded, but they also imply that distinct QFTs could interfere in a potentially destructive manner. Squeezed-state excitations and two-particle excitations imposed on the quantum vacuum seem particularly interesting~\cite{private}. In particular, the QFEC might not hold in a completely state-independent context~\cite{private}.
Work on these issues is continuing. 

\section*{Acknowledgments}
PMM acknowledges financial support from the Spanish Ministry of Education through a FECYT grant, via the postdoctoral mobility contract EX2010-0854. 
MV was supported by the Marsden Fund, and by a James Cook fellowship, both administered by the Royal Society of New Zealand.  
We wish to thank Chris Fewster, Larry Ford, and Tom Roman for their comments and feedback.

\clearpage


\end{document}